\documentclass[prd]{revtex4}

\usepackage{graphicx}
\usepackage{amsmath}
\usepackage{comment}
\usepackage{slashed}
\usepackage{CJKutf8}
\usepackage{url}
\usepackage{bm}

\usepackage{tikz}
\newcommand*{\circled}[1]{\lower.7ex\hbox{\tikz\draw (0pt, 0pt)%
		circle (.5em) node {\makebox[1em][c]{\small #1}};}}

\usepackage{tikz}
\usetikzlibrary{arrows,shapes}
\usetikzlibrary{trees}
\usetikzlibrary{matrix,arrows}
\usetikzlibrary{positioning}				
\usetikzlibrary{calc,through}				
\usetikzlibrary{decorations.pathreplacing}  

\usepackage{pgffor}							
\usetikzlibrary{decorations.pathmorphing}	
\usetikzlibrary{decorations.markings}

\tikzstyle{block} = [draw, rectangle, minimum height=3em, minimum width=6em]
\usepackage{hyperref}

\begin{document}
\pagecolor{white}
\title{Quark-quark-gluon vertex for heavy quarks up to order $1/m^5$}
\date{\today}
\author{Guojun Huang and Pengfei Zhuang}
\address{Physics Department, Tsinghua University, Beijing 100084, China}

\begin{abstract}
Instead of the often used Foldy-Wouthuysen-Tani (FWT) transformation in non-relativistic quantum chromodynamics (NRQCD), we take a more general relation between the relativistic and non-relativistic on-shell spinors to recalculate the quark-quark-gluon vertex for heavy quarks. In comparison with the previous result using FWT, the recalculated coefficients in the NRQCD Lagrangian are different at order $1/m^3$ and new at order $1/m^4$ and $1/m^5$, where $m$ is the heavy quark mass.
\end{abstract}
\maketitle

The non-relativistic quantum chromodynamics (NRQCD) is often used to describe the dynamics of heavy quarks at low energy~\cite{brambilla1,isgur,brambilla2}. The usual understanding of the total heavy quark momentum is the summation of the quark mass $m$ times its velocity $v$ and the residual momentum $k$, $p=mv+k$. Luke and Manohar constructed the transformation for spinor fields in analogy to a vector transformation, which yields the relation between the reparametrized spinor field $\Psi_v$ and the conventional heavy quark field $\psi_v$~\cite{luke},
\begin{equation}
\label{reparameter}
\Psi_v={m+i\partial\cdot v+i\slashed\partial_\perp+\sqrt{(m+i\partial\cdot v)^2+(i\partial_\perp)^2} \over \left[2\sqrt{(m+i\partial\cdot v)^2+(i\partial_\perp)^2}\left(m+i\partial\cdot v+\sqrt{(m+i\partial\cdot v)^2+(i\partial_\perp)^2}\right)\right]^{1/2}}\psi_v
\end{equation}
with the definition of a transverse vector $a_\perp^\mu = a^\mu-v^\mu a\cdot v$. On the other hand, the often used Foldy-Wouthuysen-Tani (FWT) transformation yields a simple representation~\cite{braaten}
\begin{equation}
\label{FWT1}
\Psi_v = e^{\frac{i\slashed D_\perp}{2m}}\psi_v,
\end{equation}
which can be expanded in terms of the inverse heavy quark mass,
\begin{equation}
\label{FWT2}
\Psi_v=\left(1+{\slashed k_\perp\over 2m}+{k_\perp^2\over 8m^2}+\cdots\right)\psi_v
\end{equation}
in momentum representation. To the order $1/m^2$, the two results are consistent with each other, under the condition of $i\partial\cdot v\psi_v\rightarrow k\cdot v\psi_v=0$, which is, however, not easy to be understood for $p=mv+k$. Attempts to obtain exact FWT transformation in some cases can be seen, for instance, in reference~\cite{goncalves}.

In this paper, we will recalculate the quark-quark-gluon vertex, using a more general relation for relativistic and non-relativistic on-shell fields, and fix the coefficients of the corresponding NRQCD Lagrangian by matching the result with the perturbative QCD calculation. We will also make comparison with the previous calculation using the FWT transformation to see the difference.

Under the assumption of on-shell condition $p_0=E=\sqrt{m^2+{\bf p}^2}$, which is valid for heavy quarks, the relation between the relativistic spinors $u$ and $v$ corresponding to positive and negative energies and their non-relativistic limits $u_{NR}$ and $v_{NR}$ cab be expressed as~\cite{weinberg},
\begin{eqnarray}
\label{relation}
u(p) &=& {1\over \sqrt{2E(E+m)}}(m+\slashed p)u_{NR}(p)={m+E+\slashed p_\perp\over \sqrt{2E(E+m)}}u_{NR}(p),\nonumber\\
v(p) &=& {1\over \sqrt{2E(E+m)}}(m-\slashed p)v_{NR}(p)={m+E-\slashed p_\perp\over \sqrt{2E(E+m)}}v_{NR}(p)
\end{eqnarray}
in the local rest frame with quark velocity $v_\mu=(1,0,0,0)$. The $1/m$ expansion of the relation for heavy quarks to the order $1/m^2$ is just the normal FWT transformation (\ref{FWT1}), when we take $k_\perp$ to be $p_\perp$. In this sense the relation (\ref{relation}) can be considered as a generalized FWT transformation. Under the assumption $i\partial\cdot v\psi_v\rightarrow k\cdot v\psi_v=0$ and $i\partial^\mu\psi_v\Rightarrow k^\mu\psi_v=p_\perp^\mu\psi_v$ in momentum representation, the reparameterization (\ref{reparameter}) and the relation (\ref{relation}) are also consistent to each other.

We now calculate the quark-quark-gluon vertex for heavy quarks. Taking the on-shell condition, the vertex can be expressed in terms of the form factors $F_1\left(q^2/m^2\right)$ and $F_2\left(q^2/m^2\right)$\cite{luke},
\begin{equation}
\label{qqg1}
-ig\bar{u}(p')T^a\left(\gamma^\mu F_1(q^2/m^2)+{i\sigma^{\mu\nu}q_\nu\over 2m}F_2(q^2/m^2)\right)u(p)
\end{equation}
with coupling constant $g$, Gell-Mann matrices $T_a$ and momentum transfer $q=p'-p$ between the initial and final momenta. Taking the relations $\gamma_0=\sigma_3 \otimes I_2$ and ${\bm \gamma}=i\sigma_2 \otimes {\bm \sigma}$, the vertex can be explicitly written as
\begin{eqnarray}
\label{qqg2}
&& {-ig T^a \over \sqrt{4E'(E'+m)E(E+m)}}\psi^\dag\bigg[\left(F_1(q^2/m^2)+F_2(q^2/m^2)\right)\nonumber\\
&& \times\left({\delta^\mu}_0\left((m+E')(m+E)+{\bf p}'\cdot{\bf p}+i{\bm \sigma}\cdot({\bf p}'\times{\bf p})\right)+{\delta^\mu}_j\left((m+E')\sigma^j{\bf p}\cdot{\bm \sigma}+(m+E){\bf p}'\cdot{\bm \sigma}\sigma^j\right)\right)\nonumber\\
&& -{F_2(q^2/m^2)\over 2m}(p'^\mu+p^\mu)\left((m+E')(m+E)-{\bf p}'\cdot{\bf p}-i{\bm \sigma}\cdot({\bf p}'\times{\bf p})\right)\bigg]\psi
\end{eqnarray}
with the final state energy $E'=\sqrt{m^2+{\bf p}'^2}$.

In terms of the small variable $q^2/m^2$ for heavy quarks, the form factors $F_i$ ($i=1,2$) can be expanded as
\begin{eqnarray}
\label{expansion}
F_i\left(\frac{q^2}{m^2}\right) &=& \sum_{n=0}^{+\infty}{1\over n!} {d^nF_i(q^2/m^2)\over d(q^2/m^2)^n}\Big|_{q^2/m^2=0}\left({q^2\over m^2}\right)^n\nonumber\\
&=& F_i(0)-{{\bf q}^2\over m^2}F_i'(0)+{1\over 4m^4} \left[({\bf p}^2-{\bf p}'^2)^2F_i'(0) +2{\bf q}^4F_i''(0)\right]+\cdots.
\end{eqnarray}
To simplify the notation, we will take $F_i=F_i(0), F_i'=F_i'(0)$ and $F_i''=F_i''(0)$ in the following.

Taking into account the transformation (\ref{relation}) between relativistic and non-relativistic quark fields, the expansion (\ref{expansion}) for the form factors, and the relations $\gamma_0=\sigma_3\otimes I_2$, ${\bm \gamma}=i\sigma_2\otimes{\bm \sigma}$, $u_{NR}=\left(\begin{matrix}\psi\\0\\ \end{matrix}\right)$ and $\gamma^0u_{NR}=u_{NR}$, the vertex can be expressed in terms of the current $j_\mu$,
\begin{equation}
\label{current1}
-igT^au_{NR}^\dag j_\mu A^\mu_a u_{NR}
\end{equation}
with
\begin{eqnarray}
\label{current2}
j_0 &=& F_1-{1\over 4m^2}\left[\left({1\over 2}F_1+F_2+4F_1'\right){\bf q}^2 -i\left(F_1+2F_2\right){\bm \sigma}\cdot({\bf p}'\times{\bf p})\right]\nonumber\\
&& +{1\over 8m^4}\bigg[\left({5\over 16}F_1+{1\over 4}F_2+2F_1'\right)({\bf p}^2-{\bf p}'^2)^2 +\left(F_1'+2F_2'+4F_1''\right){\bf q}^4+\left({3\over 8}F_1+{1\over 2}F_2\right)({\bf p}'^2+{\bf p}^2){\bf q}^2\nonumber\\
&&-i\left(2F_1'+4F_2'\right){\bm \sigma}\cdot({\bf p}'\times{\bf p}){\bf q}^2-i\left({3\over 4}F_1+F_2\right){\bm \sigma}\cdot({\bf p}'\times{\bf p})({\bf p}'^2+{\bf p}^2)\bigg]+\mathcal{O}(1/m^6)
\end{eqnarray}
to the order $1/m^4$ and
\begin{eqnarray}
\label{current3}
{\bf j} &=& {1\over 2m} \left[F_1({\bf p}+{\bf p}')+i\left(F_1+F_2\right){\bm \sigma}\times{\bf q}\right]-{1\over 8m^3}\bigg\{\left[F_1({\bf p}'^2+{\bf p}^2) +\left({1\over 2}F_2+4F_1'\right){\bf q}^2\right]({\bf p}+{\bf p}')\nonumber\\
&& +i\left[\left(F_1+F_2\right)({\bf p}'^2+{\bf p}^2)+4\left(F_1'+F_2'\right){\bf q}^2\right]{\bm \sigma}\times{\bf q}+{1\over 2}\left(F_1+F_2\right)({\bf p}'^2-{\bf p}^2){\bf q}\nonumber\\
&&+{i\over 2}\left(F_1+F_2\right)({\bf p}'^2-{\bf p}^2){\bm \sigma}\times({\bf p}+{\bf p}')-iF_2{\bm \sigma}\cdot({\bf p}'\times{\bf p})({\bf p}+{\bf p}')\bigg\}\nonumber\\
&&+\frac{i{\bm \sigma}\times({\bf p'+p})}{128 m^5}\left[({\bf p}'^2-{\bf p}^2){\bf p'\cdot p}\left(-16 F_1'-16 F_2'\right)+({\bf p'}^4-{\bf p}^4)\left(8 F_1'+5 F_1+8 F_2'+5 F_2\right)\right]\nonumber\\
&&+\frac{i{\bm \sigma}\times{\bf q}}{8 m^5}\bigg[({\bf p'\cdot p})^2\left(8 F_1''+8 F_2''\right)-2 ({\bf p}'^2+{\bf p}^2){\bf p'\cdot p} \left(F_1'+F_2'+4 F_1''+4 F_2''\right)\nonumber\\
&&+{\bf p}^2 {\bf p}'^2 \left(\frac{3}{16}F_1+\frac{3}{16}F_2+4 F_1''+4 F_2''\right)+\left({\bf p}'^4+{\bf p}^4\right) \left(\frac{21}{32}F_1+\frac{21}{32}F_2+2 F_1'+2 F_2'+2 F_1''+2 F_2''\right)\bigg]\nonumber\\
&&-\frac{{\bf p'+p}}{8 m^5}\left[\left(F_2'+\frac{3}{8}F_2\right) \left({\bf p}'^2+{\bf p}^2\right)-2 F_2' {\bf p'\cdot p}\right]i{\bm \sigma}\cdot({\bf p}'\times{\bf p})\nonumber\\
&&+\frac{{\bf q}}{128 m^5}\left[\left({\bf p}'^2-{\bf p}^2\right) {\bf q}^2 \left(8F_1'+8F_2'\right)+\left({\bf p}'^4-{\bf p}^4\right)(5F_1+5F_2)\right]\nonumber\\
&&+\frac{{\bf p'+p}}{16 m^5}\bigg[({\bf p'\cdot p})^2 \left(4F_2'+16F_1''\right)-\left({\bf p}'^2+{\bf p}^2\right) {\bf p'\cdot p} \left(\frac{3}{4}F_2+4 F_1'+4 F_2'+16 F_1''\right)\nonumber\\
&&+{\bf p}^2 {\bf p}'^2 \left(\frac{3}{8}F_1+\frac{1}{4}F_2+2 F_2'+8F_1''\right)+\left({\bf p}'^4+{\bf p}^4\right) \left(\frac{21}{16}F_1+\frac{5}{8}F_2+4F_1'+F_2'+4F_1''\right)\bigg]+\mathcal{O}(1/m^7)
\end{eqnarray}
to the order $1/m^5$.

We now turn to the NRQCD calculation. The corresponding NRQCD Lagrangian density can be fixed by matching the above calculated quark-quark-gluon vertex. Integrating out the unphysical field $\psi'_v$ which is defined as $\psi(x)=e^{-im v\cdot x}(h_v(x)+\psi'_v(x))$ and satisfies the relation $\slashed v\psi'_v=-\psi'_v$, the familiar Lagrangian
\begin{equation}
\mathcal L = \bar h_v\left(i D\cdot v\right)h_v+\bar h_v\left(i \slashed D_\perp{1\over 2m+iD\cdot v}i\slashed D_\perp\right) h_v
\end{equation}
can be expanded in terms of $1/m$,
\begin{eqnarray}
\mathcal L &=& \bar h_v(iD_0)h_v+{1\over 2}\sum_{n=0}^\infty{(-1)^n\over (2m)^{n+1}}\bar h_v\left\{(iD_0)^n,\ {\bf D}^2+g{\bm \sigma}\cdot{\bf B}\right\}h_v\nonumber\\
&&+{1\over 2}\sum_{n=0}^\infty{(-1)^n\over (2m)^{n+2}}\sum_{l=0}^n\bar h_v(iD_0)^{n-l}g\left(\left[{\bf D},\ {\bf E}\right]_.+i{\bm \sigma}\cdot\left[{\bf D},\ {\bf E}\right]_\times\right)(iD_0)^l h_v\nonumber\\
&&+\sum_{n=1}^\infty{(-1)^n\over (2m)^{n+2}}\sum_{l=0}^{n-1}\sum_{l'=0}^{n-l-1}\bar h_v(iD_0)^{n-1-l-l'}g^2\left({\bf E}\cdot(iD_0)^{l'}{\bf E}+i{\bm \sigma}\cdot\left({\bf E}\times(iD_0)^{l'}{\bf E}\right)\right)(iD_0)^l h_v
\end{eqnarray}
in the local rest frame with the definitions of $D_0=\partial_t-igZA_0$, ${\bf D}={\bf \nabla}-igZ{\bf A}_aT^a$, ${\bf E}_i=-G_{i0}$, ${\bf B}_i=-\epsilon_{ijk}G^{jk}/2$, $[{\bf a},\ {\bf b}]_.={\bf a}\cdot{\bf b}-{\bf b}\cdot{\bf a}$, $[{\bf a},\ {\bf b}]_\times={\bf a}\times{\bf b}-{\bf b}\times{\bf a}$ and $\{a,\ b\}_.=ab+ba$. By redefining the field
\begin{equation}
h_v\rightarrow\left[1+{{\bf D}^2+g{\bm \sigma}\cdot{\bf B}\over 8m^2}+{(iD_0)^3+g\left([{\bf D},\ {\bf E}]_. +i{\bm \sigma}\cdot[{\bf D},\ {\bf E}]_\times\right)\over 16m^3}+\cdots\right]h_v,
\end{equation}
dropping the terms unrelated to the quark-quark-gluon vertex, and taking the process similar to \cite{richard}, we have to the order $1/m^5$,
\begin{eqnarray}
\label{NRQCD2}
\mathcal L &=& h_v^\dag\bigg[ iD_0+c_2{{\bf D}^2\over 2m}+c_4{{\bf D}^4\over 8m^3}+c_6{{\bf D}^6\over 16m^5}+g {c_F {\bm \sigma}\cdot{\bf B}\over 2m} +g{c_D [{\bf D},\ {\bf E}]_.+ic_S {\bm \sigma}\cdot[{\bf D},\ {\bf E}]_\times\over 8m^2}\nonumber\\
&& +g{c_{W1} \{{\bf D}^2,\ {\bm \sigma}\cdot{\bf B}\}-2c_{W2} {\bf D}\cdot({\bm \sigma}\cdot{\bf B}){\bf D} +c_{p'p}{\color{black}(\boldsymbol{\mathrm{\sigma}}\cdot\boldsymbol{\mathrm{D}}\boldsymbol{\mathrm{B}}\cdot\boldsymbol{\mathrm{D}}+\boldsymbol{\mathrm{D}}\cdot\boldsymbol{\mathrm{B}}\boldsymbol{\mathrm{\sigma}}\cdot\boldsymbol{\mathrm{D}})}+ic_M \{{\bf D},\ {\bf B}\times{\bf D}\}_.\over 8m^3}\nonumber\\
&& +g{c_{X1}[{\bf D}^2, {\bf D}\cdot{\bf E}+{\bf E}\cdot{\bf D}]+c_{X2}\{{\bf D}^2, [{\bf D},\ {\bf E}]_.\}+c_{X3}[{\bf D}^i, [{\bf D}^i, [{\bf D},\ {\bf E}]_.]]\over m^4}\nonumber\\
&& +g{ic_{X5}{\bf D}^i{\bm \sigma}\cdot[{\bf D},\ {\bf E}]_\times{\bf D}^i+ic_{X6}\epsilon_{ijk}\sigma^i{\bf D}^j[{\bf D},\ {\bf E}]_.{\bf D}^k\over m^4}\nonumber\\
&&+g{c_{Y1}\{{\bf D}^4, {\bm \sigma}\cdot{\bf B}\}+c_{Y2}{\bf D}^2{\bm \sigma}\cdot{\bf B}{\bf D}^2+c_{Y3}\{{\bf D}^2,{\bf D}^i{\bm \sigma}\cdot{\bf B}{\bf D}^i\}+c_{Y4}{\bf D}^i{\bf D}^j{\bm \sigma}\cdot{\bf B}{\bf D}^j{\bf D}^i\over m^5}\nonumber\\
&&+g{c_{Y5}\{{\bf D}^2, {\bm \sigma}\cdot{\bf D}{\bf B}\cdot{\bf D}+{\bf D}\cdot{\bf B}{\bm \sigma}\cdot{\bf D}\}+c_{Y6}{\bf D}^i({\bm \sigma}\cdot{\bf D}{\bf B}\cdot{\bf D}+{\bf D}\cdot{\bf B}{\bm \sigma}\cdot{\bf D}){\bf D}^i\over m^5}\nonumber\\
&&+g{ic_{Y7}\{{\bf D}^2, {\bf D}\cdot({\bf B}\times{\bf D})+({\bf D}\times{\bf B})\cdot{\bf D}\}+ic_{Y8}{\bf D}^i[{\bf D}\cdot({\bf B}\times{\bf D})+({\bf D}\times{\bf B})\cdot{\bf D}]{\bf D}^i\over m^5}\nonumber\\
&&+g{c_{Y9}[{\bf D}^2,[{\bm \sigma}\cdot{\bf D}, {\bf B}\cdot{\bf D}+{\bf D}\cdot{\bf B}]]\over m^5}+\cdots\bigg]h_v
\end{eqnarray}
with the definitions of $\{{\bf a},\ {\bf b}\times{\bf c}\}_.={\bf a}\cdot({\bf b}\times{\bf c})+({\bf c}\times {\bf b})\cdot{\bf a}$, where $c_2$, $c_4$, $c_6$, $c_F$, $c_D$, $c_S$, $c_{W1}$, $c_{W2}$, $c_{p'p}$, $c_M$, $c_{X1}$, $c_{X2}$, $c_{X3}$, $c_{X5}$, $c_{X6}$, $c_{Y1}$, $c_{Y2}$, $c_{Y3}$, $c_{Y4}$, $c_{Y5}$, $c_{Y6}$, $c_{Y7}$, $c_{Y8}$ and $c_{Y9}$ are the coefficients to be determined. The first four terms in the Lagrangian determine the quark propagator in the gauge field $A_\mu^a$ as well as contribute to the quark-quark-gluon vertex, and the other terms control the quark-quark-gluon vertex~\cite{manohar,dye}.

It is direct to take out the quark-quark-gluon vertex from the NRQCD Lagrangian density (\ref{NRQCD2}). By matching it with the current (\ref{current1}) controlled by the form factors, we extract the NRQCD coefficients to the order $1/m^5$,
\begin{eqnarray}
\label{coefficient1}
c_2 &=& c_4=c_6=1,\nonumber\\
c_F &=& F_1+F_2,\nonumber\\
c_D &=& F_1+2F_2+8F_1',\nonumber\\
c_S &=& F_1+2F_2,\nonumber\\
c_{W1} &=& F_1+F_2/2+4F_1'+4F_2',\nonumber\\
c_{W2} &=& F_2/2+4F_1'+4F_2',\nonumber\\
c_{p'p} &=& F_2,\nonumber\\
c_M &=&F_2/2+4F_1',\nonumber\\
c_{X1} &=& {5F_1/128}+{F_2/32}+{F'_1/4},\nonumber\\
c_{X2} &=& 3F_1/64+F_2/16,\nonumber\\
c_{X3} &=& F'_1/8+F_2'/4+F_1''/2,\nonumber\\
c_{X5} &=& 3F_1/32+F_2/8,\nonumber\\
c_{X6} &=& -3F_1/32-F_2/8-F_1'/4-F_2'/2,\nonumber\\
c_{Y1} &=& 27F_1/256+23F_2/256+5F_1'/16+5F_2'/16+F_1''/4+F_2''/4,\nonumber\\
c_{Y2} &=& -3F_1/128-11F_2/128-F_1'/8-3F_2'/8+F_1''/2+F_2''/2,\nonumber\\
c_{Y3} &=& -3F_2/64-F_1'/4-F_2'/4-F_1''-F_2'',\nonumber\\
c_{Y4} &=& F_2'/4+F_1''+F_2'',\nonumber\\
c_{Y5} &=& 3F_2/64+F_2'/8,\nonumber\\
c_{Y6} &=& -F_2'/4,\nonumber\\
c_{Y7} &=& 3F_2/128+F_1'/8+F_2'/16+F_1''/4,\nonumber\\
c_{Y8} &=& -F_2'/8-F_1''/2,\nonumber\\
c_{Y9} &=& -3F_1/128-F_2/32-F_1'/16-F_2'/8.
\end{eqnarray}

We now compare our result with the previous calculations. Instead of the general transformation (\ref{relation}) between relativistic and non-relativistic on-shell spinors which is employed in the above calculation, the usual FWT transformation (\ref{FWT1}) was used in the previous calculations. To the order $1/m^3$ the result here is consistent to the previous work~\cite{manohar}, except for the coefficient $c_M$, which is consistent to the $Variational\ method$ result in NRQED~\cite{NRQED_to_1overm4}, while the coefficients of order $1/m^4$ are consistent to the NRQED results~\cite{NRQED_to_1overm4}. Firstly, the coefficient $c_M$ at order $1/m^3$ are different in two transformation. Secondly, the previous calculation is only up to the order $1/m^3$, but the calculation here can be up to the $1/m^5$ and higher order which results in the coefficients $c_{X1}, c_{X2}, c_{X3}, c_{X5}$, $c_{X6}$, $c_{Y1}$, $c_{Y2}$, $c_{Y3}$, $c_{Y4}$, $c_{Y5}$, $c_{Y6}$, $c_{Y7}$, $c_{Y8}$ and $c_{Y9}$.

The final step to determine the coefficients is to calculate the form factors $F_1(q^2/m^2)$ and $F_2(q^2/m^2)$ in QCD at some specific level~\cite{gerlach}. For instance, computing the Feynman diagrams to one-loop correction and taking into account the related renormalization, we have
\begin{eqnarray}
F_1\left({q^2\over m^2}\right) &=& 1+{\alpha_s\over 144\pi}{q^2\over m^2}\left[\left(-51+154\ln {m\over \mu}\right) +{1\over 10}{q^2\over m^2}\left(131+888 \ln{m\over \mu}\right)\right],\nonumber\\
F_2\left({q^2\over m^2}\right) &=& {\alpha_s\over 6\pi}\left\{\left(13-9\ln {m\over \mu}\right)+{q^2\over m^2}\left[{1\over 6}\left(13-54 \ln{m\over \mu}\right)-{3\over 4}{q^2\over m^2}\left(1+6 \ln {m\over \mu}\right)\right]\right\}
\end{eqnarray}
with the redefined coupling constant $\alpha_s=g^2/(4\pi)$ and the cut-off $\mu$ in dimensional renormalization.

With the known form factors it is straightforward to represent the coefficients with $m$ and $\mu$,
\begin{eqnarray}
\label{coefficient3}
c_F &=& 1+\frac{\alpha_s}{6 \pi }\left(13-9 \ln\frac{m}{\mu}\right),\nonumber\\
c_D &=& 1+\frac{\alpha_s}{18 \pi } \left(27+100 \ln\frac{m}{\mu }\right),\nonumber\\
c_S &=& 1+\frac{\alpha_s}{3 \pi }\left(13-9 \ln\frac{m}{\mu}\right),\nonumber\\
c_{W1} &=& 1+\frac{\alpha_s}{36 \pi }\left(40-89\ln\frac{m}{\mu }\right),\nonumber\\
c_{W2} &=& \frac{\alpha_s}{36 \pi }\left(40-89 \ln\frac{m}{\mu}\right),\nonumber\\
c_{p'p} &=& \frac{\alpha_s}{6 \pi }\left(13-9 \ln\frac{m}{\mu}\right),\nonumber\\
c_{M} &=&{\alpha_s\over 36\pi}\left(-12+127\ln{m\over\mu}\right),\nonumber\\
c_{X1} &=&{5\over 128}+{\alpha_s\over 576\pi}\left(-12+127\ln{m\over\mu}\right),\nonumber\\
c_{X2} &=&{3\over 64}+{\alpha_s\over 96\pi}\left(13-9\ln{m\over\mu}\right),\nonumber\\
c_{X3} &=&{\alpha_s\over 5760\pi}\left(789+2162\ln{m\over\mu}\right),\nonumber\\
c_{X5} &=&{3\over 32}+{\alpha_s\over 48\pi}\left(13-9\ln{m\over\mu}\right),\nonumber\\
c_{X6} &=&-{3\over 32}+{\alpha_s\over 576\pi}\left(-209+386\ln{m\over\mu}\right),\nonumber\\
c_{Y1} &=&{27\over 256}+{\alpha_s\over 23040\pi}\left(4143-7741\ln{m\over\mu}\right),\nonumber\\
c_{Y2} &=&-{3\over 128}+{\alpha_s\over 11520\pi}\left(-3587+4889\ln{m\over\mu}\right),\nonumber\\
c_{Y3} &=&{\alpha_s\over 5760\pi}\left(-203+2561\ln{m\over\mu}\right),\nonumber\\
c_{Y4} &=& {\alpha_s\over 360\pi}\left(8-231\ln{m\over\mu}\right),\nonumber\\
c_{Y5} &=& {\alpha_s\over 1152\pi}\left(169-297\ln{m\over\mu}\right),\nonumber\\
c_{Y6} &=& {\alpha_s\over 144\pi}\left(-13+54\ln{m\over\mu}\right),\nonumber\\
c_{Y7} &=& {\alpha_s\over 11520\pi}\left(859+3607\ln{m\over\mu}\right),\nonumber\\
c_{Y8} &=& {\alpha_s\over 720\pi}\left(-98-309\ln{m\over\mu}\right),\nonumber\\
c_{Y9} &=& -{3\over 128}+{\alpha_s\over 2304\pi}\left(-209+386\ln{m\over\mu}\right).
\end{eqnarray}

In summary, we recalculated the quark-quark-gluon vertex for heavy quarks. Instead of the usual FWT transformation which is often used in previous calculations, we employed the more general relation between relativistic and non-relativistic on-shell spinors. By matching our calculation with the standard NRQCD calculation, the coefficients in the NRQCD Langrangian are fixed. While the result to the order $1/m^2$ is the same with the previous calculation using the FWT transformation, some coefficients at order $1/m^3$ are already different, and those coefficients at orders $1/m^4$ and $1/m^5$ which can not be determined in the pervious calculation are derived in the current calculation.
\\ \\
{\bf Acknowledgement}: We thank Mr. Jiaxing Zhao and Yan-qing Ma for helpful discussions during the work. The work is supported by the NSFC grant No. 11890712.

\end{document}